\documentstyle[12pt]{article}
\begin{document}

\baselineskip=18pt
\begin{tabbing}
\hskip 12.5 cm \= {SISSA-46/95/EP}\\
\hskip 1 cm \>{May 1995}\\
\end{tabbing}
\thispagestyle{empty}
\vskip 0.3cm
\begin{center}
{\bf SUPERSYMMETRIC RADIATIVE CORRECTIONS}
\vskip 0.3cm
{\bf TO NEUTRINO INDICES OF REFRACTION}
\vskip 1.0 cm
{\bf Esteban Roulet}
\vskip .3cm
SISSA, Trieste, 34100, Italy

\vskip 1. cm
{ABSTRACT}
\end{center}

\noindent We compute the one-loop effects on the neutrino propagation
through matter induced by virtual supersymmetric particles. We show
that, in the minimal version of the supersymmetric standard model, a
non-degeneracy between sleptons of the second and third
generations can have sizeable effects on the $\nu_\mu$--$\nu_\tau$
oscillations in matter. In particular, we discuss how this could
affect the detection of the energetic neutrino fluxes arising from
annihilation of supersymmetric dark matter in the center of the sun.

\vskip 0.5 cm

\newpage
\baselineskip=18pt

The fact that  $W$-boson exchange with electrons in ordinary matter
affects the $\nu_e$ index of refraction but not those of the other
neutrino flavours is the basis of the Mikheyev-Smirnov-Wolfenstein
(MSW) effect~\cite{msw}, which provides the nicest explanation to the solar
neutrino problem.

In the Standard Model (SM), at small momentum transfer muon and tau
neutrinos interact
indistinguishably with ordinary matter at the tree level. However, a
difference among the $\nu_\mu$ and $\nu_\tau$ indices of refraction in
matter appears at one-loop~\cite{bo87}, although it is suppressed by
$O\left({\alpha\over \pi \sin^2\theta_W}{m_\tau^2\over M_W^2}\right)$
with respect to the size of the charged current effects affecting the
$\nu_e$ propagation, having then probably no observable implications.

In this paper we compute the one-loop contributions to the neutrino
refraction indices in the best motivated extension of the SM, i.e. the
minimal supersymmetric version of it (MSSM). We show that there are
potentially much larger radiative effects than in the SM itself and we
discuss the possible physical relevance they could have.

The interactions of neutrinos with matter are described by the matrix
element
\begin{equation}M(\nu_\ell f\to \nu_\ell f)=-i{G_F\over
\sqrt{2}}\bar\nu_\ell\gamma^\rho(1-\gamma_5)\nu_\ell\bar f\gamma_\rho
(C^V_{\nu_\ell f}+C^A_{\nu_\ell f}\gamma_5)f .
\end{equation}
For neutrinos propagating through an unpolarized medium at rest, the
temporal component of the fermionic vector current leads
to a non-vanishing neutrino forward-scattering amplitude and, hence, to a
neutrino refraction index $n_\nu$ given by~\cite{la83}
\begin{equation}
p_\nu(n_{\nu_\ell}-1)=-\sqrt{2}G_F\sum_{f=u,d,e}C^V_{\nu_\ell f}N_f
.\end{equation}
$N_f$ is the number density of fermion $f$ and, at the tree-level,
\begin{equation}C^V_{\nu_\ell f}=T_3(f_L)-2Q_f s_W^2+\delta_{\ell f} ,
\end{equation}
with $s_W^2\equiv \sin^2\theta_W$, $T_3(f_L)$ the third component of
isospin of $f_L$ and $Q_f$ its charge.

The indices of refraction affect the neutrino flavour evolution during
propagation, which is described by (for reviews see~\cite{review})
\begin{equation}i{d\over dt}\left(\begin{array}{r}\nu_e\\ \nu_\mu\\
\nu_\tau\end{array}\right)=
\left[ {1\over 2p_\nu}V\left(\begin{array}{ccc}\Delta m^2_{12}&0&0\\0&0&0\\
0&0&\Delta m^2_{32}\end{array}\right) V^\dagger
-p_\nu\left(\begin{array}{ccc}
\Delta n_{e \mu}&0&0\\0&0&0\\0&0&\Delta n_{\tau\mu}\end{array}\right)\right]
\left(\begin{array}{r}\nu_e\\ \nu_\mu\\ \nu_\tau\end{array}\right),
\end{equation}
where $V$ is the unitary matrix relating the neutrino flavour
($\nu_\alpha$, with $\alpha=e,\ \mu,\ \tau$) and mass ($\nu_i$, with
$i=1,$ 2, 3) eigenstates, i.e. $\nu_\alpha=V_{\alpha i}\nu_i$. Also
$\Delta m^2_{ij}\equiv m_{\nu_i}^2-m_{\nu_j}^2$ and $\Delta
n_{\alpha\beta}\equiv n_{\nu_\alpha}-n_{\nu_\beta}$. (We note that,
although the indices of refraction will be computed in the MSSM, some
departure from it should be responsible for the neutrino masses and
mixings themselves).

Since matter effects in oscillations involving $\nu_e$ will be largely
dominated by the charged current (CC) term $\delta_{\ell f}$ in eq. (3),
the radiative corrections to the tree-level result
$\Delta n_{e\mu}=-\sqrt{2}G_FN_e/p_\nu$ turn out to be
negligible. We will then
concentrate on the computation of $\Delta n_{\tau\mu}$, which does not
involve the CC piece and vanishes at the tree-level.

It is useful to parametrise the effects
of the radiative corrections, for $f\neq \ell$, as
\begin{equation}C^V_{\nu_\ell f}=\rho^{\nu_\ell
f}T_3(f_L)-2Q_f\lambda^{\nu_\ell f}s_W^2
.\end{equation}
The splitting of the
radiative effects between $\rho$ and $\lambda$ is somewhat arbitrary,
and it is convenient to include in $\rho$ the $f$-dependent box diagram
contributions. This has the advantage that, for a neutral
medium, the corrections in $\lambda$ (which include also contributions
from the neutrino charge radius~\cite{chra}) turn out to be $f$-independent
(see ref.~\cite{bo87}) and do not
contribute to $\Delta n_{\tau\mu}$, due to the fact that
\begin{equation}\sum_{f=u,d,e}N_fQ_f=0 .\end{equation}

One has then
\begin{equation}p_\nu\Delta
n_{\tau\mu}=-\sqrt{2}G_F\sum_fN_fT_3(f_L)\Delta\rho^f
,\end{equation}
with $\Delta\rho^f\equiv \rho^{\nu_\tau f}-\rho^{\nu_\mu f}$.

In the SM, $\Delta\rho^f$ gets contributions from the one-loop
corrections to the $\bar\nu\nu Z$ vertex and $W$-boson box diagrams
\cite{bo87},
leading to
\begin{equation}\Delta\rho^e_{SM}=\Delta\rho^d_{SM}={\alpha_W\over
8\pi }\left[{x(2+x)\over
1-x}+{3x(2-x)\over (1-x)^2}{\rm ln}x\right]\end{equation}
\begin{equation}\Delta\rho^u_{SM}={-\alpha_W\over 8\pi }\left[{x(4-x)\over
1-x}+{3x^2\over (1-x)^2}{\rm ln}x\right] ,\end{equation}
with $x\equiv m_\tau^2/M_W^2$ and $\alpha_W\equiv\alpha/s_W^2$.
These corrections are small due to the
one-loop factor $\alpha_W/4 \pi $, but are also quite suppressed by
the smallness of the factor $m_\tau^2/M_W^2\sim 4\times 10^{-4}$.

The computation of the supersymmetric contribution to $\Delta\rho$
requires the evaluation of the Feynman diagrams depicted in fig.~1,
leading to the results that are summarized in the Appendix. The
important point is that, besides the one-loop factor $\alpha_W/4\pi $,
what fixes now the size of $\Delta\rho$ is mainly the splitting among
the sleptons of the second and third generations.

At this point, it is useful to recall that a usual simplifying
assumption made in phenomenological applications is to consider all
sfermions to be exactly degenerate at the GUT scale, and obtain their
low energy splittings from the renormalization group evolution of the
soft parameters and from terms arising after the electroweak symmetry
breaking. In this way, although
squarks get significantly splitted from sleptons,
the splittings among the masses of different slepton generations
are only due to the small $\tau$-Yukawa coupling. This usually implies
that
$m_{\tilde \tau}^2-m_{\tilde\mu}^2$ is $O(m_\tau^2)$, and hence
the radiative effects on the $\nu_{\mu,\tau}$ indices of
refraction are in this case not larger than the SM ones. An
exception to this, still assuming a universal soft scalar mass $m$,
is when there is a large $\tilde\tau_L$--$\tilde\tau_R$ mixing.
This happens for large
values of the Higgs mixing parameter $\mu$ and large tg$\beta$, or for
large values of the parameter $A$ of the trilinear soft terms, in
which case the splitting can be  $O[m_\tau(Am+\mu{\rm
tg}\beta$)] (see ref.~\cite{haber}).

{}From a more general perspective, the universality
assumptions (which give the easiest way to get rid of FCNC phenomena) are
not really a necessity, and actually non-universal soft terms usually
arise in string theories~\cite{string} and can also be generated
in GUTs~\cite{po95}.
Universal sfermion masses may not even be desirable in some
respects, and it has been argued that non-universalities
may prove useful in reconciling different phenomenological
constraints in supersymmetric GUTs~\cite{ol94,po95}.
 Also, it has recently been suggested that sfermion masses may
dynamically align along the directions, in flavour space, of the
fermion masses, suppressing FCNC but allowing large mass splittings
\cite{di95}.

If one considers the general case in which a sizeable
splitting is allowed among $\mu$ and $\tau$
sleptons\footnote{This splitting is not directly related to
very suppressed rare processes such as $\mu\to e\gamma$, $\mu\to 3e$,
$\tau\to\mu\gamma$, etc.~\cite{ch95},
but could give rise to small universality
violations~\cite{ch94}.}, the SUSY contribution
to $\Delta n_{\tau\mu}$ could then be larger than the SM
one.

We will present an illustrative situation in which the effect here
described is important and then comment on how the results are
modified when one changes the starting assumptions. For simplicity we
assume no $\tilde f_L$--$\tilde f_R$ mixings, neglect
intergenerational mixings of sleptons as well as splittings due to
$D$-terms among charged and neutral sleptons or among $\tilde\ell_L$
and $\tilde\ell_R$, which are anyhow inessential to the conclusions
reached. We take first generation sleptons degenerate with the second
generation ones, and only allow the third generation sleptons to have
a different mass. We take a light but experimentally allowed value for
the second generation slepton masses,
$m_{\tilde\mu}=m_{\tilde\nu_\mu}={\rm Max}[$60~GeV, $m_\chi+20$~GeV],
where $m_\chi$ is the lightest neutralino mass (which we assume to be
the lightest supersymmetric particle (LSP)), and take second and third
generation sleptons to be splitted by an amount
$m_{\tilde\tau}-m_{\tilde\mu}=60$~GeV. We also assume that squarks are
much heavier than sleptons (as usually results from the effects of
gluino masses in the renormalization group evolution of scalar
masses). This last implies that box diagrams only contribute sizeably
to interactions with electrons.

In fig.~2 we plot the ratio of the SUSY and SM values of $\Delta
n_{\tau\mu}$ for an isoscalar medium ($Y_n\equiv N_n/N_p=1$), as a
function of the supersymmetric parameter space that determines the
chargino and neutralino masses and couplings (SU(2) gaugino
mass\footnote{for definiteness we assumed common gaugino masses at the
GUT scale to obtain the neutralino spectra} $M$
and Higgs mixing parameter $\mu$). We present results for values of
the ratio of Higgs VEVs tg$\beta\equiv v_2/v_1=2$ (fig.~2.$a$) and
tg$\beta=40$ (fig.~2.$b$), showing that the dependence on it is only
mild. The dark regions for
small values of $\mu$ and $M$ are
excluded by the LEP constraint $m_{\chi^+}>45$~GeV, that is the main
bound from accelerators.

It is apparent from fig.~2
that, for the slepton mass splittings considered,
$\Delta n_{\tau\mu}$ may be an order of magnitude larger
than in the SM. The SUSY contribution turns out to be dominated
by the chargino boxes and penguins involving $\tilde \ell_L$ exchange.
The neutralino boxes are generally small, while  neutralino penguins,
not shown in fig.~1, give no contribution (similarly to what happens
for instance in $b\to s\ell^+\ell^-$ decays~\cite{be91}).
Thus, the relevant
splitting is the one among charged sleptons rather than among
sneutrinos. The effect is especially large in a region of parameter
space where the chargino masses are below 80--100~GeV, i.e. testable
at LEPII, and becomes less important for large values of
$|\mu|$ and $M$, i.e. for heavier charginos. Slepton
splittings smaller than the one adopted
would lead to proportionally smaller effects, while larger
splittings can increase the effect by up to a factor of two.
The neglected box diagrams involving squark exchange may
also increase the SUSY contribution.
 Also
note that, since penguin contributions to $\Delta\rho$ are
$f$-independent, they lead (see eq. (7)) to a contribution to
$\Delta n_{\tau\mu}$ proportional to $N_e(2-Y_n)$ (in the sun,
$Y_n$ varies from $\sim 0.16$ in the surface to 0.5 in the center).

Clearly the sign of $\Delta n_{\tau\mu}$, and hence whether
resonant matter effects take place among neutrinos or
antineutrinos\footnote{for antineutrinos, the
sign of the matrix element is
reversed, so that $n_{\bar\nu_\tau}-n_{\bar\nu_\mu}=-(
n_{\nu_\tau}-n_{\nu_\mu})$}, depends, for significant slepton
splittings, on whether ${\tilde\tau}$ are heavier or lighter than
$\tilde\mu$ ($m_{\tilde\tau}$ smaller than $m_{\tilde\mu}$ leads to a
resonance crossing among neutrinos if $\Delta m^2_{32}>0$).

Let us also note that the penguin and box diagrams involving $\tilde\ell_R$
exchange are proportional to the square of the lepton Yukawa coupling,
so that their effect on $n_{\nu_\mu}$ is much smaller than
that on $n_{\nu_\tau}$. This fact has the interesting effect
of making their contribution to $\Delta n_{\tau\mu}$ to depend just on
$m_{\tilde\tau}$, rather than on a slepton mass splitting. However,
even for tg$\beta=40$ and $m_{\tilde\tau_R}=60$~GeV, they give a
contribution not larger than the SM one. Similar conclusions hold for
the penguins involving $H^+$ exchange (keeping in mind that in the
MSSM $m_{H^+}>M_W$).

We finally mention that other extensions of the SM may also sizeably
affect $\Delta n_{\tau\mu}$.
In particular, supersymmetric $R$-parity violating
interactions can modify the neutrino indices of refraction already  at
the tree-level~\cite{ro91}, although those models would be less
interesting as regards the application discussed below. Another simple
example would be the presence of a new $Z'$ gauge boson with
non-universal couplings to leptons~\cite{na93}.

We turn now to consider the possible physical relevance of these
radiative effects for $\nu_\mu$--$\nu_\tau$ matter oscillations. When
discussing applications, we will neglect the $\nu_{\mu,\tau}$ mixing
with $\nu_e$  to be left with just a two flavour situation.
The generalization to three flavour neutrino mixing should
pose no problems.

A first difficulty to observe any conversion among $\nu_\mu$ and
$\nu_\tau$ is that for low energies, $E_\nu<0.1$~GeV, these neutrinos
are only detected by means of their neutral current interactions and
are hence essentially
indistinguishable. Furthermore, only $e$-type neutrinos are
produced in the sun (except for possible ordinary MSW conversions inside the
sun) and equal amounts of $\nu_\mu$ and $\nu_\tau$ are produced in
supernovae, so that oscillations among them do not give actually any
overall result. These problems are not present in long-baseline
$\nu_\mu$ oscillation experiments on earth (either with $\nu_\mu$ from
accelerators or using atmospheric neutrinos), but it is easy to
convince oneself that the resonance oscillation length in terrestrial
matter (inversely proportional to $\Delta n_{\tau\mu}$)
 is typically much larger than the earth diameter, and hence
oscillation effects are negligible.

The situation that we want to describe, in which the matter effects
here analysed are relevant, is actually directly related to the
supersymmetric framework under consideration. It is well known that a
nice feature of supersymmetry, once $R$-parity conservation is adopted
to avoid $B$ and $L$ violation, is that the LSP, usually a neutralino,
is stable and naturally becomes a good dark matter (DM) candidate. There
are two main strategies that are being pursued at present to
experimentally search for SUSY DM~\cite{dm}.
The first is the direct search of
the energy deposited by halo neutralinos interacting with target nuclei,
and the second one is the search of energetic neutrinos produced in
the annihilation of DM trapped in the interior of the sun or the
earth~\cite{search}\footnote{halo $\chi$ annihilations may also provide some
signals.}. In particular, upward-going muons produced in the rock (or
ice) just below underground detectors by energetic $\nu_\mu$ and
$\bar\nu_\mu$ (with $E_\nu>$~few GeV) may allow to probe significant
regions of the supersymmetric parameter space in new
installations such as Superkamiokande or Amanda. As was shown in ref.
\cite{el92}, the usual MSW effect between $\nu_e$ and $\nu_{\mu,\tau}$
can affect the detection rate predictions.

To show the possible effects of energetic $\nu_\mu$--$\nu_\tau$ matter
enhanced oscillations in the solar interior, we plot in fig.~3 the
$\nu_\mu$ neutrino survival probability in the
sin$^22\theta_{\mu\tau}$ vs. $\Delta m^2_{32}$ plane, assuming that
$\epsilon\equiv\Delta n_{\tau\mu}/\Delta n_{\mu e}=10^{-3}$
(in the SM, $\epsilon \simeq -5\times 10^{-5}$). The contours correspond
to $P(\nu_\mu\to\nu_\mu)=0.8$ (continuous lines) and 0.45 (dashed
lines) for two neutrino energies\footnote{for
energies above 100~GeV, $\nu$ absorption in the sun starts to be
relevant, making the oscillation formalism to be no longer valid. Our
computation of the box diagrams, which neglected external momenta,
actually contained terms proportional to $p_\nu\cdot p_f$, which would
modify the results only for $E_\nu>$ few TeV.}, $E_\nu=10$ and 40~GeV.
It is clear  that, for significant ranges of $\Delta
m^2$ and $\sin^22\theta$, the oscillations of high energy neutrinos are
sizeably affected by matter effects.
For decreasing values of $|\epsilon|$, the MSW type resonant effects
take place for smaller  $\Delta m^2$ values. The adiabatic
condition in the resonance transition becomes harder to achieve,
making the regions of significant transition to shrink towards
large mixings, and the effect eventually becomes very small for
$|\epsilon|<10^{-4}$.

These $\nu_\mu$--$\nu_\tau$ oscillations may have important
implications for the detection of neutralino annihilation signals.
This is because the $\nu_\mu$ and $\nu_\tau$ fluxes from neutralino
annihilation are generally quite different, so that oscillations among
them modify the expected $\nu_\mu$ signal at underground detectors.
The difference among $\nu_\mu$ and $\nu_\tau$ fluxes has its origin
in the fact that
the non-relativistic neutralino
annihilation cross section into fermion pairs $f\bar f$
is proportional to $m_f^2$, either due to a $p$-wave suppression
(as
for annihilations mediated by $Z$ or sfermion exchange) or due to a
Yukawa suppression (as in the case of Higgs boson mediated
annihilations). Hence, neutralinos do not directly
annihilate into neutrino pairs. Different neutrino fluxes result then from
$b$, $c$ and $\tau$ decays~\cite{ritz} (for $m_\chi<M_W$, since otherwise other
channels involving gauge bosons in the final state are also allowed and can
produce prompt secondary neutrinos of different flavours in similar
amounts).
Furthermore, light mesons and muons produced in the $\chi$ annihilation
are stopped by the solar medium before they decay, yielding no
secondary fluxes of energetic neutrinos.

Rather than scanning all the supersymmetric parameter space,
we will consider as an illustrative example the simple but still quite
general case in which the
lightest neutralino is mainly gaugino, i.e. $|\mu|> M$, with
$m_\chi<M_W$ (in this region the effect here discussed is potentially
large and also the neutralino cosmological relic density is usually
significant).
 If squarks are heavier than sleptons, as we are
assuming, and sleptons are
not too heavy, the main non-relativistic neutralino annihilation channel
is by $t$-channel $\tilde\tau$ exchange, producing a $\tau\bar\tau$
pair. The $\nu_\tau$ and $\nu_\mu$ fluxes from the subsequent $\tau$
decays will then clearly be quite
different. (In the general case of an arbitrary neutralino composition
and squark masses, the fluxes are still different but one needs to
include the extra annihilation channels and the model dependent
branching ratios entering in the $\nu_{\mu,\tau}$ yields).

In fig.~4 we show the $\nu_\tau$ and $\nu_\mu$ differential neutrino yields
(with thin dashed and solid lines respectively)
produced by the annihilation of neutralinos into a $\tau$ pair (the
main annihilation channel in our example).
What is actually plotted is $z^2dN/dz$ (where $z\equiv
E_\nu/m_\chi$), which is the relevant quantity for underground
$\nu_\mu$ detection\footnote{the muon flux arising from $\nu_\tau$
interactions is quite suppressed}
 because both the CC $\nu_\mu$ cross section and
the muon range in the rock (or ice) are proportional to the neutrino
energy.

With thick lines we show how the fluxes get modified after
traversing the solar interior, assuming $m_\chi=50$~GeV and taking
($\epsilon,\Delta m^2,\sin^22\theta$) to be ($10^{-3}$,
$6\times 10^{-4}$~eV$^2$, 0.1) in fig.~4.$a$,  ($10^{-3}$,
$3\times 10^{-4}$~eV$^2$, 0.1) in fig.~4.$b$  and ($10^{-3}$,
$10^{-3}$~eV$^2$, 0.6) in fig.~4.$c$.
The neutrino masses and mixings assumed in the case of fig.~4.$c$
lie in the region, also shown in fig.~3,
of interest to explain the atmospheric neutrino anomaly. One should
note, however, that for large mixings the effects of vacuum oscillations
are significant and would affect both neutrinos and antineutrinos,
while matter induced oscillations affect either one or the other.

{}From the results, one can see that the detection rates may be sizeably
modified by the matter enhanced oscillations here described,
providing an interesting  physical manifestation of the radiative
supersymmetric effects studied.
The uncertainties
involved in the theoretical predictions of the DM annihilation signal
(unknown neutralino mass and composition, uncertainties in the local
halo density and DM velocity distribution, etc.) will however complicate the
interpretation of any positive detection, so that these effects
should actually be considered as providing a further spread in the
theoretical predictions until these parameters become more constrained
by accelerators and direct DM searches.
The shape of the neutrino spectrum gives probably a clearer signature
of the matter oscillations,
and may provide a useful handle to identify them.

\bigskip
\bigskip
I would like to thank Stefano Bertolini for very useful discussions.

\bigskip
{\large\bf Appendix}

We summarize here the supersymmetric contributions to
$\Delta n_{\tau\mu}$ arising from the diagrams in fig.~1.

A few simplifying assumptions will be adopted. Motivated by the
smallness of FCNC phenomena, we will assume that sfermion masses
align, in flavour space, in the directions of the fermion masses. This
makes the neutralino--fermion--sfermion vertices diagonal in
generation space.
We will also ignore $\tilde f_L$--$\tilde f_R$ mixings,
so that the appropriate mass eigenstates are
$\tilde f_{iL}$ and $\tilde f_{iR}$.
These assumptions simplify the
calculations but are not essential to the conclusions reached.

Following the notation of ref.~\cite{haber},
we denote by $Z'_{ij}$ the $4\times
4$ matrix diagonalizing the neutralino states in the basis $(\tilde
\gamma, \tilde Z,\tilde H_1, \tilde H_2)$, and $U$ and $V$ are the
$2\times 2$ matrices required for the diagonalization of the chargino
mass matrix.
In the radiative corrections involving the neutralinos $\chi^0_i$,
only the gaugino
components will contribute sizeably since the higgsino couplings are
very small for interactions with ordinary matter. The Feynman rule for
the $\bar f\chi^0_j\tilde f_L$ vertex can be then parametrised as
$-ig\sqrt{2}G^{j*}_{fL}P_R$ while that involving $\tilde f_R$ by
$ig\sqrt{2}G^{j*}_{fR}P_L$. The couplings $G^j$ to the gaugino components
are given by
$$G^j_{fL}=Q_fs_WZ^{\prime *}_{j1}+{c_L^f\over c_W}Z^{\prime *}_{j2}$$
$$G^j_{fR}={\rm sign}(m_{\chi^0_j})\left[ Q_fs_WZ'_{j1}+{c_R^f\over
c_W}Z'_{j2}\right]$$
with $c^f_{L(R)}=T_3(f_{L(R)})-Q_fs_W^2$.

Regarding the chargino couplings, we will ignore Cabibbo-type
intergenerational mixings (for leptons\footnote{this may not be a good
approximation for neutrino mixings close to maximal}
and, in the box contributions
also for quarks), since the flavour-conserving processes under study
are not GIM suppressed. We then only include the chargino mixing
matrices in the vertices (e.g. the $\bar\nu_\ell\chi^+_j\tilde\ell_L$
vertex is $-igU_{j1}P_R$, the $\bar\nu_\ell\chi^+_j\tilde\ell_R$
vertex is $igH^\ell_jP_R$, with $H^\ell_j\equiv
m_\ell U_{j2}/\sqrt{2}M_Wc\beta$, see ref.~\cite{haber}).

The penguin and box contributions to $\Delta\rho^f$ can be written as
$$\Delta\rho^f=\Delta\rho_p+{\Delta\rho^f_{box}\over T_3(f_L)}.$$
The ($f$-independent)
contributions to the penguins involving charginos and $\tilde\ell_L$
exchange, arise from the self-energy
diagrams in fig.~1.$a$, $\Delta\rho^L_p(\Sigma)$, and from the diagrams
where the $Z$ couples to the slepton-line $\Delta\rho^L_p(\tilde\ell)$
(fig.~1.$b$),
or to the chargino-line, $\Delta\rho^L_p(\chi^+)$ (fig.~1.$c$).
Direct computation leads to the results:
$$\Delta\rho^L_p(\Sigma)={\alpha_W\over 8\pi}
\sum_{j=1}^2\vert U_{j1}\vert^2\left\{
G_0(X_{\chi^+_j\tilde\tau_L},1)+{\rm ln}
{m_{\tilde\tau_L}^2\over \mu^2}-(\tilde\tau_L\to \tilde\mu_L)
\right\}$$
$$\Delta\rho^L_p(\tilde \ell)=-{\alpha_W\over 4\pi }c^\ell_L
\sum_{j=1}^2\vert U_{j1}\vert^2\left\{
G_0(X_{\chi^+_j\tilde\tau_L},1)+{\rm ln}
{m_{\tilde\tau_L}^2\over \mu^2}-(\tilde\tau_L\to \tilde\mu_L)
\right\}$$
\begin{eqnarray*}
\Delta\rho^L_p(\chi^+)&=&-{\alpha_W\over 4\pi }
\sum_{i,j=1}^2U_{i1}U^*_{j1}
\left\{2{\cal O'}^L_{ij}
\sqrt{X_{\chi^+_i\tilde\tau_L}X_{\chi^+_j\tilde\tau_L}}
F_0(X_{\chi^+_i\tilde\tau_L},X_{\chi^+_j\tilde\tau_L} )-\right.\\
&\ &\left.
{\cal
O'}^R_{ij}\left[G_0(X_{\chi^+_i\tilde\tau_L},X_{\chi^+_j\tilde\tau_L})+{\rm
ln}{m_{\tilde\tau_L}^2\over\mu^2}\right]-(\tilde\tau_L\to \tilde\mu_L)
\right\},\end{eqnarray*}
where $X_{ab}\equiv(m_a/m_b)^2$ and $\mu$ is an arbitrary mass scale.
The couplings
${\cal O'}^L_{ij}=-V_{i1}V^*_{j1}-\frac{1}{2}V_{i2}V^*_{j2}+\delta_{ij}s_W^2$
and ${\cal O'}^R_{ij}=-U_{i1}U^*_{j1}-\frac{1}{2}U_{i2}U^*_{j2}+
\delta_{ij}s_W^2$ describe the $\chi^+_i\chi^+_jZ$ vertex, that reads
$i\frac{g}{c_W}\gamma^\mu[{\cal O'}^L_{ij}P_L+{\cal O'}^R_{ij}P_R]$.
The functions $F_0$ and $G_0$ are
$$F_0(x,y)={x{\rm ln}x\over (x-y)(x-1)}+(x\leftrightarrow y),$$
$$G_0(x,y)=\left[{x^2{\rm ln}x\over (x-y)(x-1)}+(x\leftrightarrow y)\right]
-\frac{3}{2}.$$

The penguins involving $\tilde\ell_R$ exchange, that although
proportional to $(m_\tau/M_W)^2$ may in principle be enhanced for large
values of
tg$\beta$, give a total contribution (neglecting the $\tilde\mu_R$
exchange that is proportional to $m_\mu^2/M_W^2$)
\begin{eqnarray*}
\Delta\rho^R_p&\simeq&{\alpha_W\over 4\pi}\sum_{j=1}^2
\left\{\vert H^\tau_j\vert^2 \left({1\over 2}-s_W^2\right)
G_0(X_{\chi^+_j\tilde\tau_R},1)
+\sum_{i=1}^2H^\tau_iH_j^{\tau *}\left[{\cal O'}^R_{ij}\times
\right.\right.\\
&\ &\left.\left.
\left[
G_0(X_{\chi^+_i\tilde\tau_R},X_{\chi^+_j\tilde\tau_R})+1\right]-
2{\cal O'}^L_{ij}\sqrt{
X_{\chi^+_i\tilde\tau_R}X_{\chi^+_j\tilde\tau_R}}
F_0(X_{\chi^+_i\tilde\tau_R},X_{\chi^+_j\tilde\tau_R} )
\right]\right\}.
\end{eqnarray*}
Similarly, the charged Higgs boson contribution in fig.~1.$a$--$c$ is
$$\Delta\rho_p^{H^+}\simeq{\alpha_W\over 4\pi}{m_\tau^2\over
M_W^2}{\rm tg}^2\beta{y\over 2}\left[{1\over 1-y}+{{\rm ln}y\over
(1-y)^2}\right],$$
where $y\equiv (m_\tau/M_{H^+})^2$.

The box diagrams (fig.~1.$d$) involving charginos and $\tilde\ell_L$
exchange give:
\begin{eqnarray*}
\Delta\rho^e_{box}(\chi^+)&=&-{\alpha_W\over 4\pi }{M_W^2\over
m_{\tilde\nu_e}^2}\sum_{j,k=1}^2V^*_{j1}V_{k1}U^*_{j1}U_{k1} \times \\
&\ & \left\{\sqrt{X_{\chi^+_j\tilde\nu_e}
X_{\chi^+_k\tilde\nu_e}}
F'(X_{\chi^+_j\tilde\nu_e}, X_{\chi^+_k
\tilde\nu_e},
X_{\tilde\tau_L\tilde\nu_e})-(\tilde\tau_L\to\tilde\mu_L)\right\},
\end{eqnarray*}
$$\Delta\rho^d_{box}(\chi^+)=\Delta\rho^e_{box}(\chi^+)\
(\tilde\nu_e\to\tilde u_L),$$
$$\Delta\rho^u_{box}(\chi^+)=
{\alpha_W\over 8\pi }{M_W^2\over
m_{\tilde d_L}^2}\sum_{j,k=1}^2
U_{j1}U^*_{k1}U^*_{j1}U_{k1}\left\{G'(X_{\chi^+_j\tilde d_L}, X_{\chi^+_k
\tilde d_L},
X_{\tilde\tau_L\tilde d_L})-(\tilde\tau_L\to\tilde\mu_L)\right\}.$$
The boxes with $\tilde\ell_R$ exchange  are just obtained by replacing
in the previous expressions $\tilde\ell_L\to\tilde\ell_R$,
$U^*_{j1}U_{k1}\to H^\tau_jH^{\tau *}_k$ and omitting the
$\tilde\mu_R$ exchange contribution.

Finally, box diagrams involving neutralinos (fig.~1.$e,f$) interacting with
$f=e,\ u,\ d$ give:
\begin{eqnarray*}\Delta\rho^f_{box}(\chi^0)&=&-
{\alpha_W\over \pi }\sum_{j,k=1}^4
G^k_{\nu L}G^{*j}_{\nu L}\left\{{M_W^2\over m_{\tilde f_L}^2}\left[
G^k_{fL}G^{*j}_{fL}\sqrt{X_{\chi^0_j\tilde f_L}
X_{\chi^0_k\tilde f_L}}F'(X_{\tilde\nu_\tau\tilde f_L}, X_{\chi^0_j\tilde
f_L}, X_{\chi^0_k\tilde f_L})-\right.\right.\\
&\ &\left. \left. {G^j_{fL}G^{*k}_{fL}\over 2}
G'(X_{\tilde\nu_\tau\tilde
f_L}, X_{\chi^0_j\tilde f_L}, X_{\chi^0_k\tilde f_L})\right]+
(j\leftrightarrow k,\ L\leftrightarrow R)\right\}-(\tilde\nu_\tau\to
\tilde\nu_\mu),\end{eqnarray*}
where
$$F'(x,y,z)=-{x{\rm ln}x\over (x-y)(x-z)(x-1)}-
{y{\rm ln}y\over (y-x)(y-z)(y-1)}
-{z{\rm ln}z\over (z-x)(z-y)(z-1)},$$
$$G'(x,y,z)={x^2{\rm ln}x\over (x-y)(x-z)(x-1)}+
{y^2{\rm ln}y\over (y-x)(y-z)(y-1)}
+{z^2{\rm ln}z\over (z-x)(z-y)(z-1)}.$$

%\pagebreak

%\pagebreak
\noindent{\bf Figure Captions}

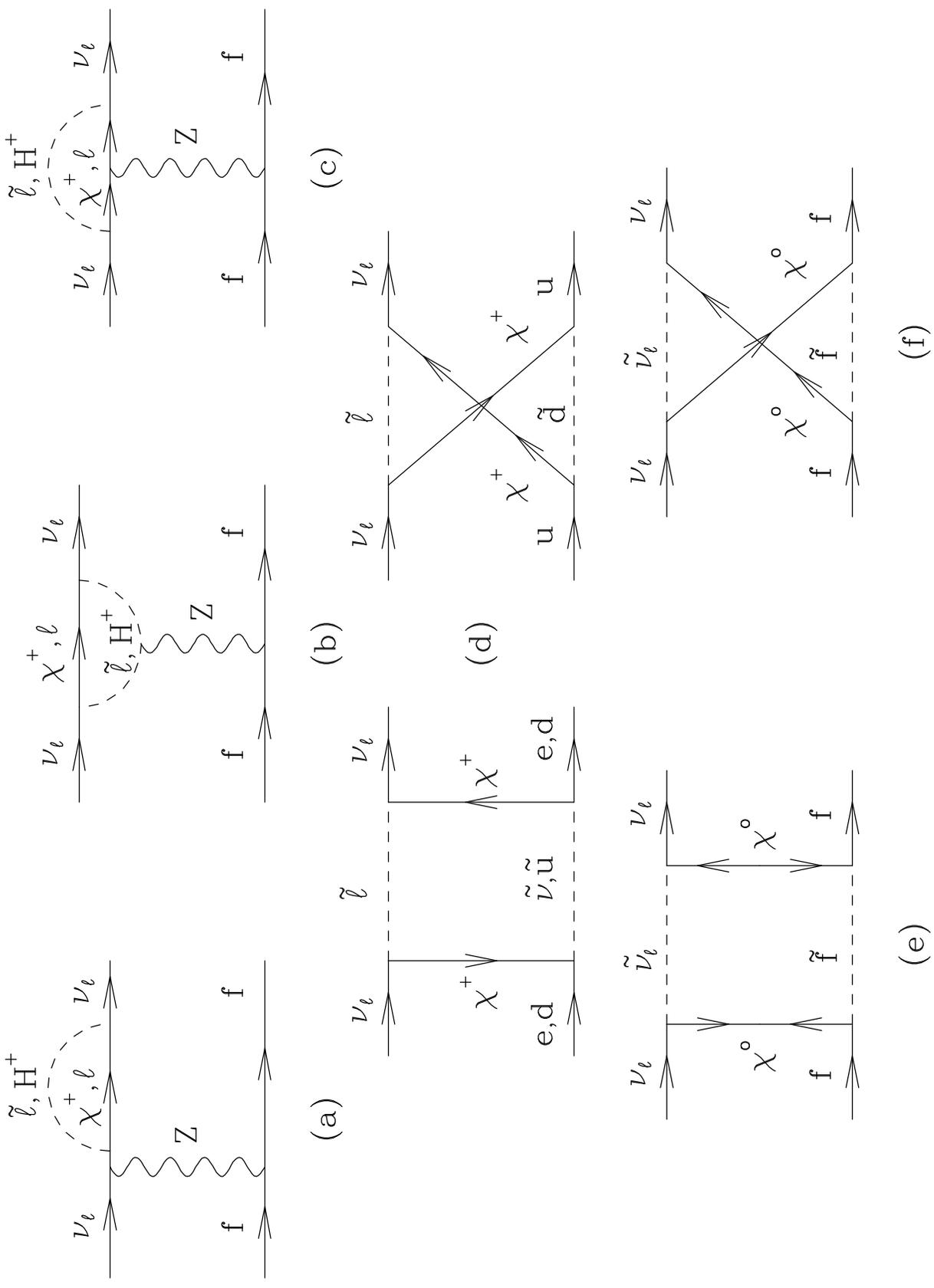
\noindent Fig.~1: Feynman diagrams describing the supersymmetric
contribution to $\nu_\ell$--$f$ forward scattering ($\ell=\mu,\
\tau$). The blob in fig.~1.$a$ represents corrections to both $\nu$
external legs.

\noindent Fig.~2: Ratio of the supersymmetric and SM values of $\Delta
n_{\tau\mu}$ for an isoscalar medium. We
take the sleptons of the first two generations to have a common mass
of Max[60~GeV, $m_\chi+20$~GeV] (see text), and assume the third
slepton generation to be heavier by 60~GeV. Fig.~2.$a$ is for
tg$\beta=2$ while fig.~2.$b$ is for tg$\beta=40$.

\noindent Fig.~3: Contours of survival probability
$P(\nu_\mu\to\nu_\mu)=0.8$ (solid lines) and 0.45 (dashed lines)
for neutrino energies $E_\nu=10$ and 40~GeV, taking
$\epsilon\equiv\Delta n_{\tau\mu}/\Delta n_{\mu e}=10^{-3}$.

\noindent Fig.~4: Differential $\nu_\mu$ (solid lines) and $\nu_\tau$
(dashed lines) yields ($\times z^2$) vs. $z\equiv E_\nu/m_\chi$, for
$\chi$ annihilations into $\tau$ pairs.
Thin lines describe the original spectra (ref.~\cite{ritz}) while
thick lines include the  matter effects, assuming $m_\chi=50$~GeV.
The figures correspond to
($\epsilon,\Delta m^2,\sin^22\theta$) equal to $(10^{-3}$,
$6\times 10^{-4}$~eV$^2$, 0.1)  in fig.~4.$a$,  ($10^{-3}$,
$3\times 10^{-4}$~eV$^2$, 0.1) in fig.~4.$b$  and ($10^{-3}$,
$10^{-3}$~eV$^2$, 0.6) in fig.~4.$c$.

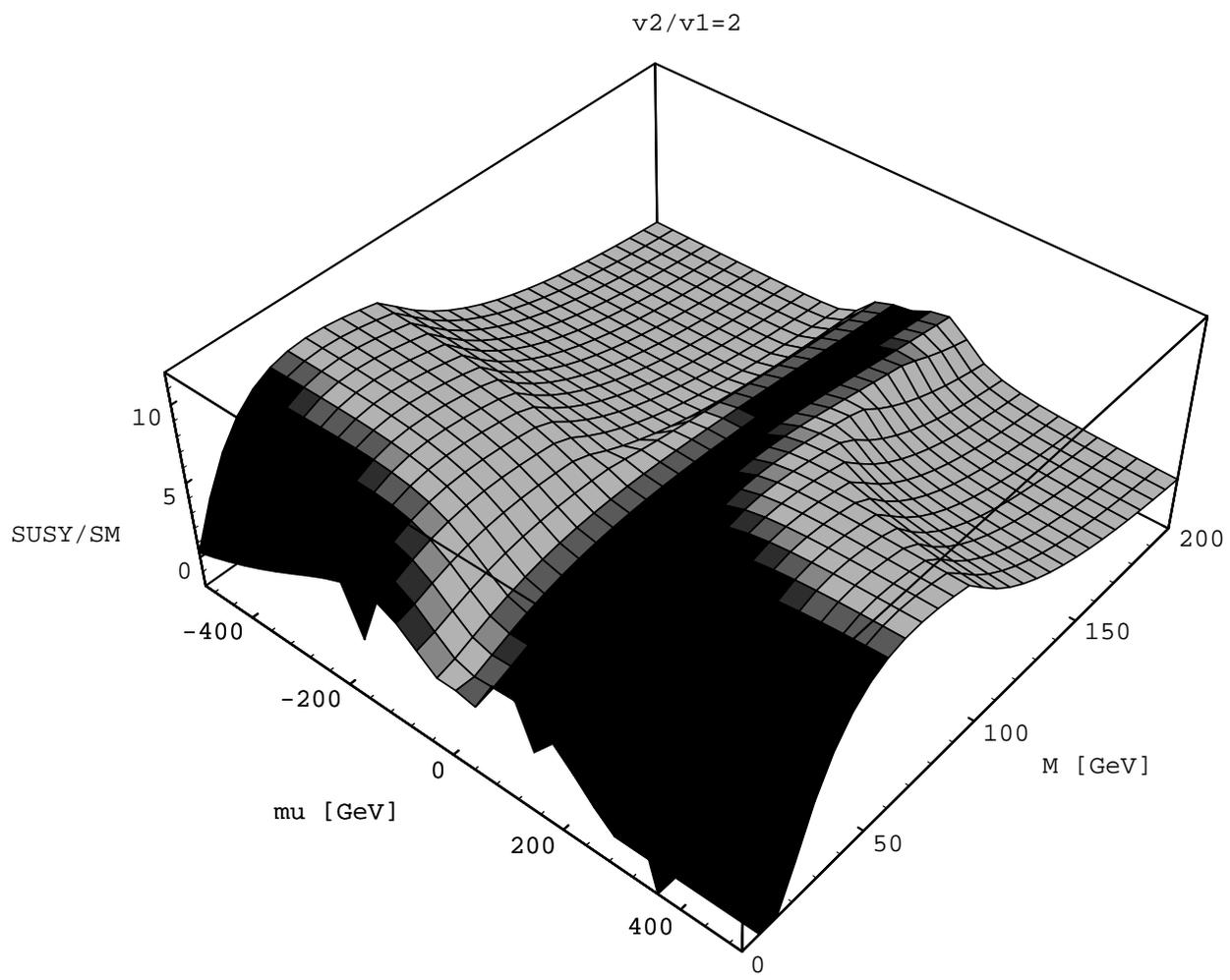
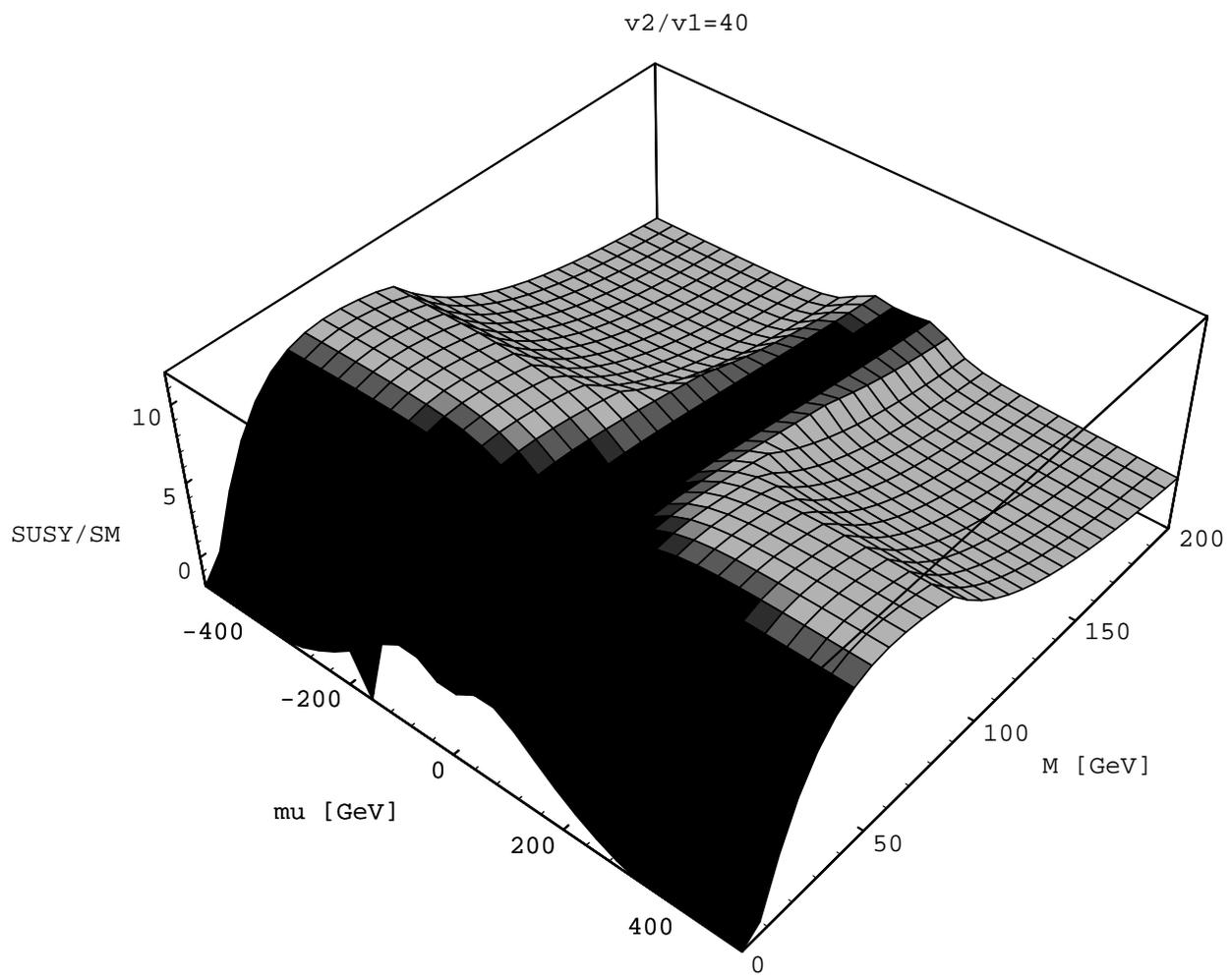
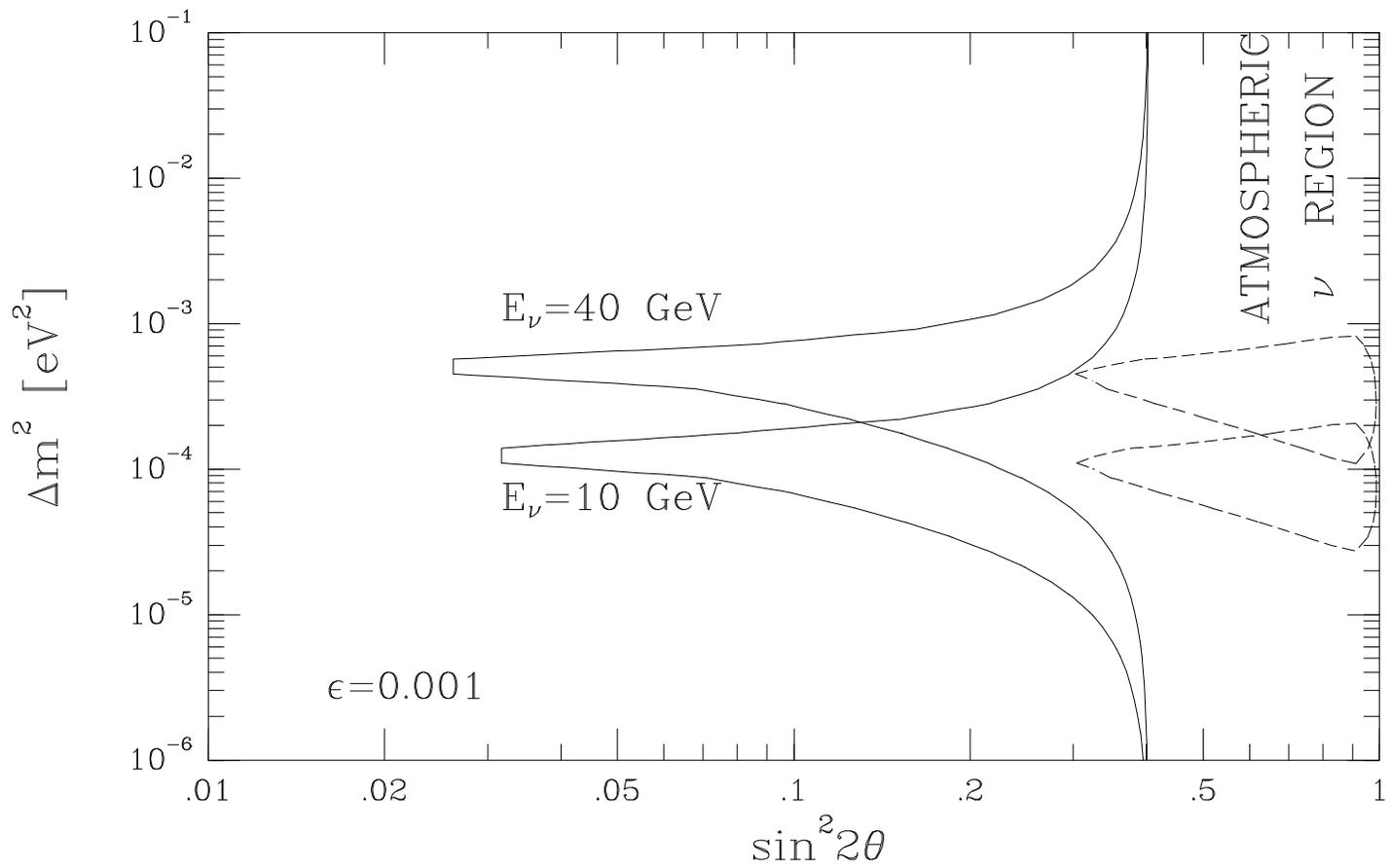
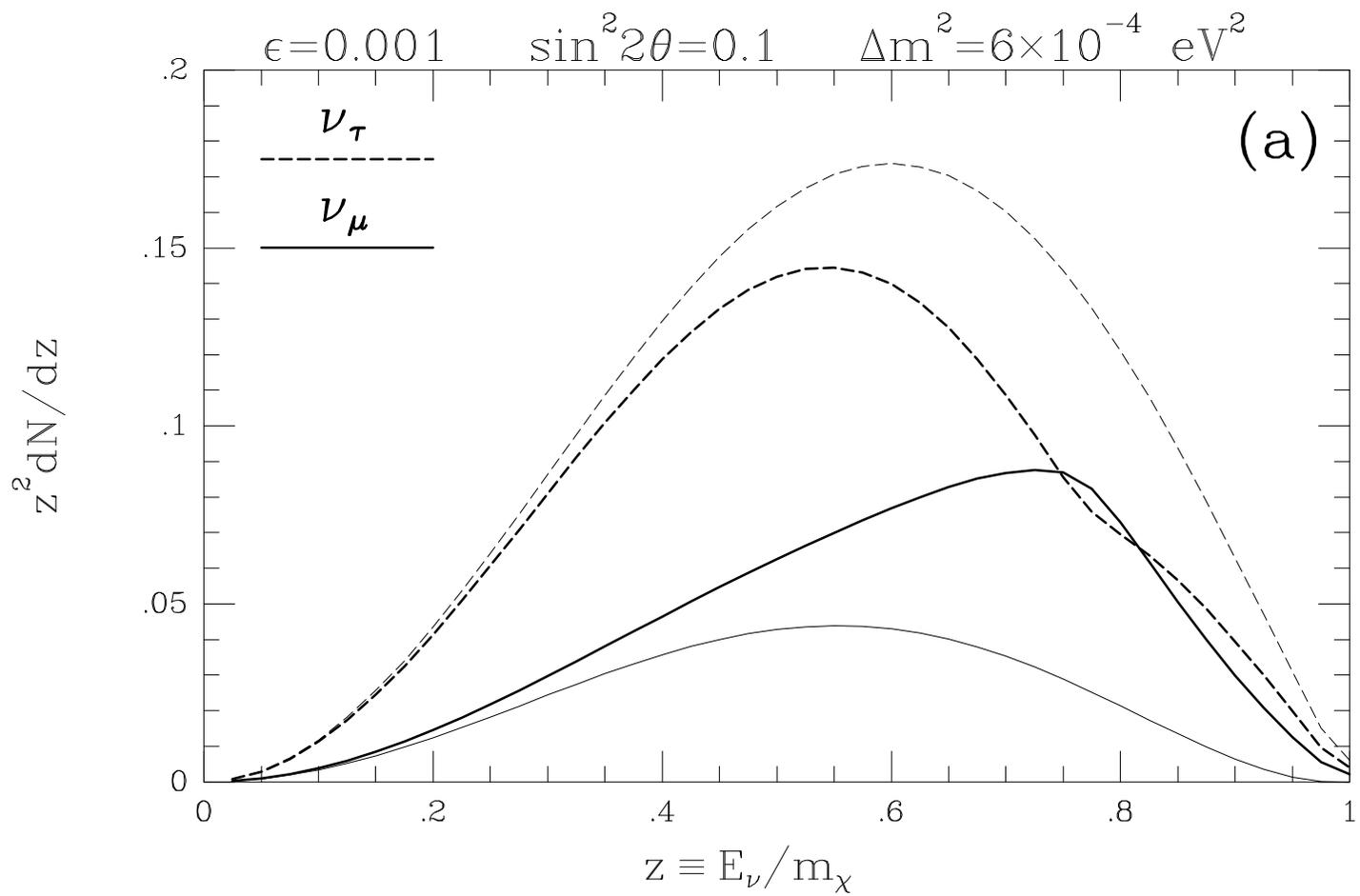
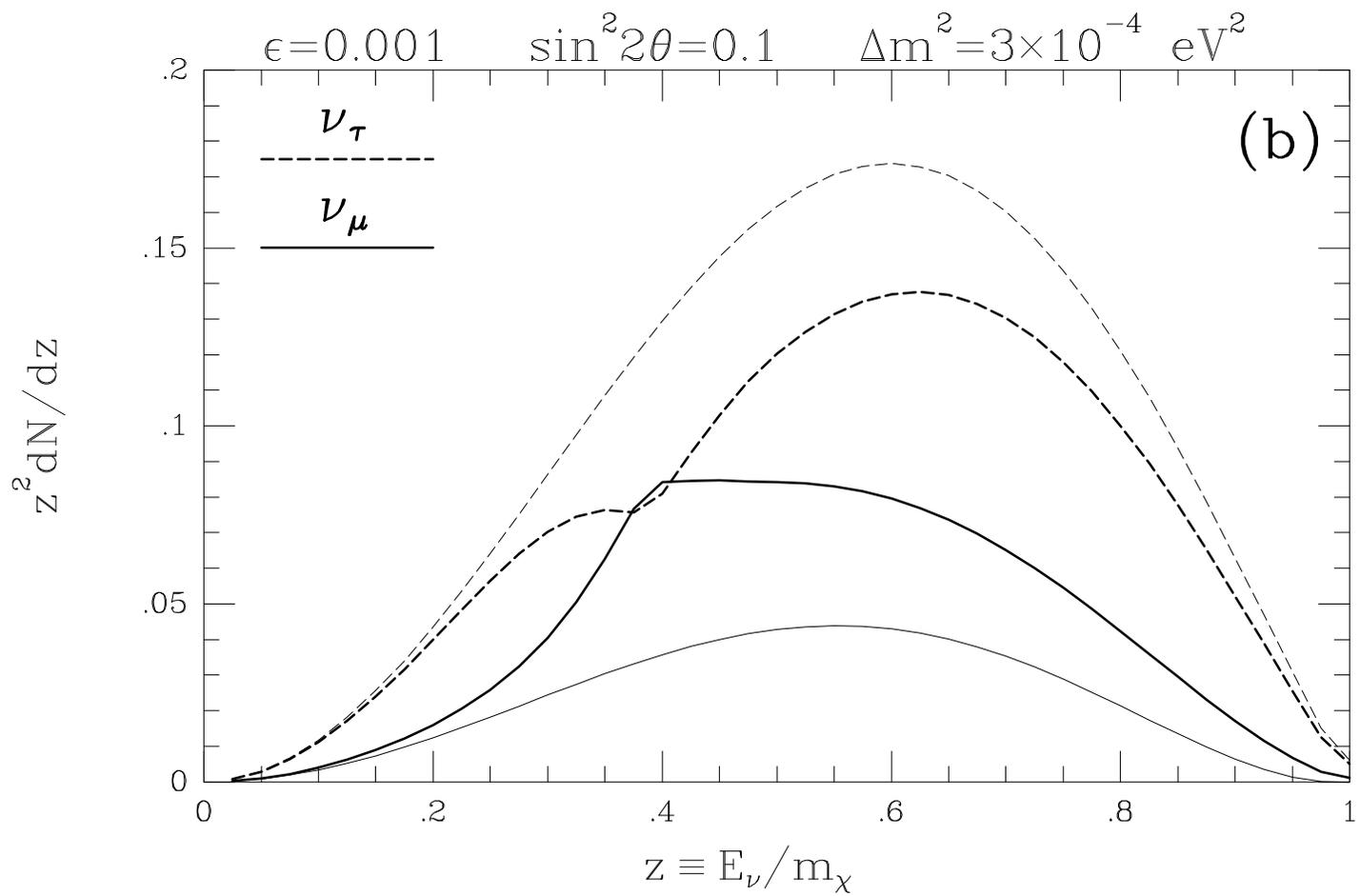
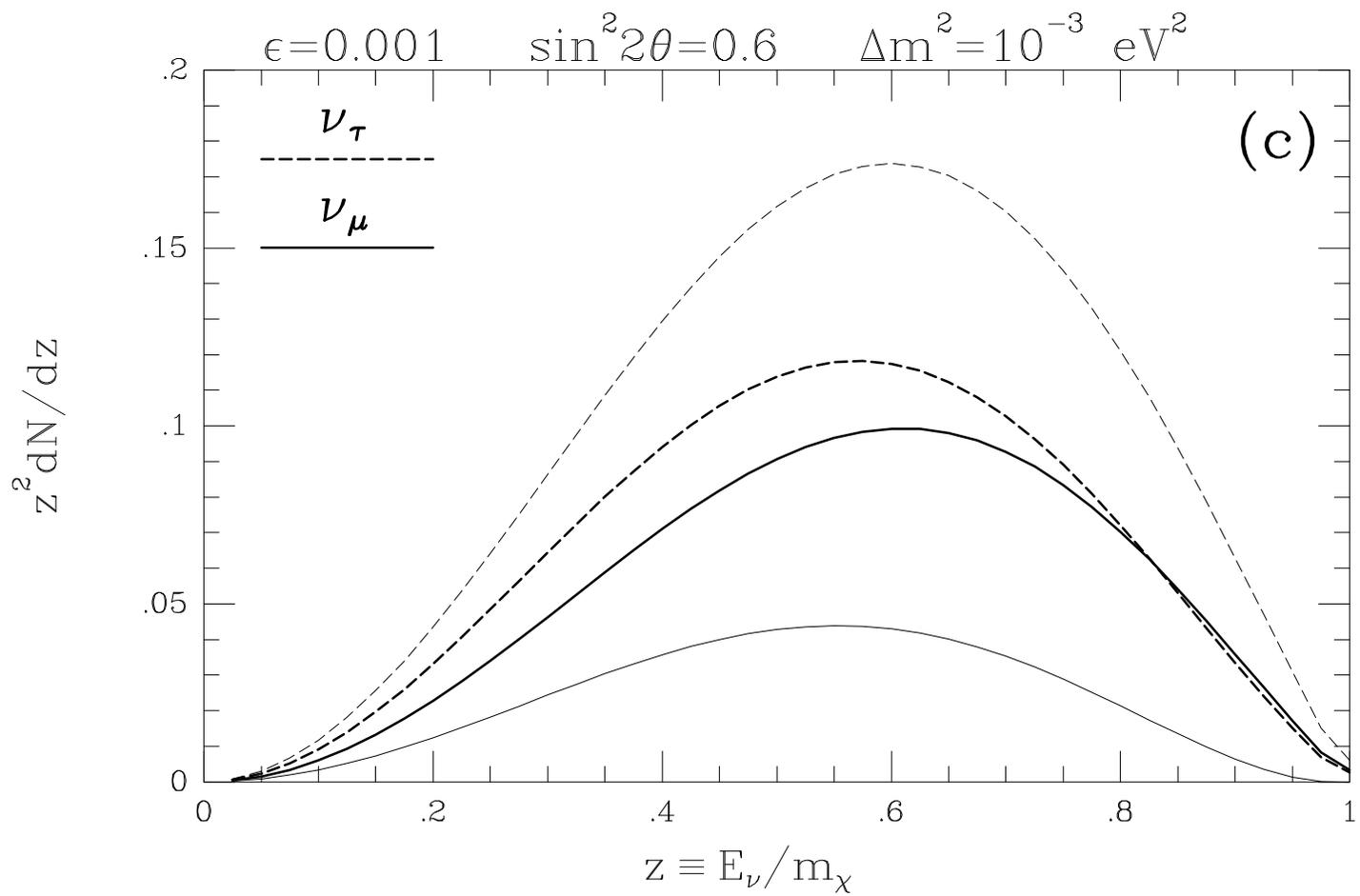
\end{document}